\documentclass[12pt,a4paper]{article}
\usepackage{amsfonts}
\usepackage{amssymb}
\usepackage{amsmath}
\usepackage{latexsym}
\begin{document}

\begin{center}
{\Large \bf Submanifolds in five-dimensional pseudo-Euclidean
\\ \vspace{2mm} spaces and four-dimensional FRW universes} \\

\vspace{4mm}

Igor E. Gulamov$\,^1$, Mikhail N. Smolyakov$\,^2$\\
\vspace{0.5cm} $^1$Physics Department, Moscow State University,
119991, Moscow, Russia\\
$^2$Skobeltsyn Institute of Nuclear Physics, Moscow State
University,\\ 119991 Moscow, Russia\\
\end{center}

\begin{abstract}
Equations for submanifolds, which correspond to embeddings of the
four-dimensional FRW universes in five-dimensional
pseudo-Euclidean spaces, are presented in convenient form in
general case. Several specific examples are considered.
\end{abstract}

It is well known that in general case a four-dimensional
pseudo-Riemannian manifold can be represented as a submanifold in
a flat ten-dimensional space-time \cite{Eddington}. If the metric
possesses additional symmetries the dimensionality of the ambient
space-time may be smaller, for example, it is well known that the
de Sitter space $dS_{4}$ can be represented as a hyperboloid in
five-dimensional Minkowski space (for the first time explicit
formulas for this embedding were presented in
\cite{Robertson1929}). It is less known that the FRW
(Friedmann-Robertson-Walker) space-times can also be embedded in
five-dimensional Minkowski space (more generally, in
pseudo-Euclidean spaces as we will show below). For the first time
this fact was noticed almost 80 years ago in \cite{Robertson},
where explicit formulas for the embeddings of the FRW universes
were presented (see note D in \cite{Robertson}). In 1965 analogous
formulas were presented in \cite{Rosen}, where various embeddings
of the solutions to equations of General Relativity in
pseudo-Euclidean spaces were considered. Nevertheless, in spite of
the recent activity in the field of embeddings of the
four-dimensional General Relativity in five-dimensional flat or
Ricci-flat space-times (see, for example, \cite{Wesson1,Wesson2}
and references therein), it looks as if the results of
\cite{Robertson,Rosen} on the embedding of the FRW space-times are
not widely known (though they can be found in textbook
\cite{Robertson-Noonan}, pp. 413--414), contrary to the case of
the de Sitter space, which is discussed in many textbooks devoted
to gravitation and cosmology (see, for example,
\cite{Weinberg,Misner,Rindler}). Indeed, formulas for embeddings
of the FRW space-times in five-dimensional Minkowski space were
reinvented in \cite{Lachieze-Rey} (for the open, closed and
spatially-flat FRW universes) and in \cite{Smolyakov} (for the
spatially-flat FRW universe). But although the explicit formulas
of embeddings for the FRW universes in five-dimensional Minkowski
space are known, we have failed to find equations for
submanifolds, which correspond to such embeddings, in explicit and
convenient form. In this note we will derive the known formulas
for embeddings of the open and closed FRW universes (taking into
account two possible embeddings for the open FRW universe) and
obtain the corresponding equations for submanifolds. For
completeness we will also present the equation for the case of the
spatially-flat FRW universe. Finally, we will consider several
specific examples.

First, we consider the case of the open universe. Let us take a
five-dimensional space-time with the flat metric
\begin{equation}\label{metric}
ds^{2}=-dt'^{2}+d{\vec y}^{2}\pm
d\eta^2=-dt'^2+dr^2+r^2\left(d\theta^2+\sin^2(\theta)d\varphi^2\right)\pm
d\eta^2.
\end{equation}
Note that the extra dimension with coordinate $\eta$ can be
space-like or time-like. With the help of transformations
\begin{eqnarray}\label{M1}
r=\tau\sinh(\chi),\\ \label{M2} t'=\tau\cosh(\chi)
\end{eqnarray}
one can obtain the metric
\begin{equation}
ds^2=-d\tau^2+\tau^2\left(d\chi^2+\sinh^2(\chi)\left(d\theta^2+\sin^2(\theta)d\varphi^2\right)\right)\pm
d\eta^2,
\end{equation}
which resembles the metric of the Milne universe \cite{Milne} (for
details see also \cite{Rindler}, pp. 204--207; and "the expanding
Minkowski universe" in \cite{Robertson-Noonan}, pp. 362--365 and
in \cite{Misner}, pp. 743--744) apart from the term $d\eta^2$. Now
it is easy to find the appropriate submanifold for the case of the
open FRW universe. Let us suppose that the coordinates on the
hypersurface are $\chi$, $\theta$, $\varphi$ and $t$, such that
\begin{equation}\label{M3}
\tau=a(t),\quad \eta=b(t).
\end{equation}
We get
\begin{equation}\label{frw1}
ds^2=-\left({\dot a}^{2}(t)\mp {\dot
b}^{2}(t)\right)dt^2+a^{2}(t)\left(d\chi^2+\sinh^2(\chi)\left(d\theta^2+\sin^2(\theta)d\varphi^2\right)\right),
\end{equation}
where $\dot{}=\frac{d}{dt}$. Thus,
\begin{eqnarray}\label{b1}
\textrm{if}\quad & & {\dot a}^{2}(t)>1,\quad {\dot b}^{2}(t)={\dot
a}^{2}(t)-1\quad \textrm{(space-like extra dimension)},\\
\label{b2} \textrm{and if}\quad & & {\dot a}^{2}(t)<1,\quad {\dot
b}^{2}(t)=1-{\dot a}^{2}(t)\quad \textrm{(time-like extra
dimension)},
\end{eqnarray}
we get the four-dimensional FRW metric of the open universe.
Equations (\ref{M1}), (\ref{M2}), (\ref{M3}) with (\ref{b1}) can
be found in \cite{Rosen,Lachieze-Rey}. The case (\ref{b2}) was
mentioned in \cite{Robertson} and considered in
\cite{Robertson-Noonan} as an embedding of the open universe. We
will show below that it leads to some peculiar consequences and is
rather unphysical in the standard FRW cosmology, contrary to the
case (\ref{b1}).

The equation of the appropriate submanifold can be easily
obtained. Indeed, using (\ref{M1}), (\ref{M2}) and (\ref{M3}) we
get
\begin{equation}\label{paramet1}
t'^2-r^2=\left.a^2(t)\right|_{t=b^{-1}(\eta)}.
\end{equation}
The latter formula can be rewritten in another form. Let us
consider the Friedmann equation (here and below we consider the
standard FRW cosmology)
\begin{equation}\label{Friedmann}
\left(\frac{\dot a}{a}\right)^2=\frac{8\pi
G}{3}\rho-\frac{\kappa}{a^2},
\end{equation}
$\kappa=-1$ for the open universe. Using this equation one can
obtain from (\ref{M3}), (\ref{b1}) and (\ref{b2})
\begin{eqnarray}
\eta=\pm\int dt\sqrt{\frac{8\pi G}{3}\rho a^2}=\pm\int
da\sqrt{\frac{\frac{8\pi G}{3}\rho a^2}{1+\frac{8\pi G}{3}\rho
a^2}},\\ \nonumber \textrm{($\rho>0$, space-like extra
dimension)},
\end{eqnarray}
\begin{eqnarray}
\eta=\pm\int dt\sqrt{\frac{8\pi G}{3}|\rho| a^2}=\pm\int
da\sqrt{\frac{\frac{8\pi G}{3}|\rho| a^2}{1-\frac{8\pi G}{3}|\rho|
a^2}},\\ \nonumber \textrm{($\rho<0$, time-like extra dimension)},
\end{eqnarray}
which leads to the following equations for the hypersurfaces:
\begin{eqnarray}\label{op_pos}
\eta=\pm\left.\int\sqrt{\left(\frac{\frac{8\pi G}{3}\rho
a^2}{1+\frac{8\pi G}{3}\rho
a^2}\right)}da\right|_{a=\sqrt{t'^2-r^2}},\\ \nonumber
\textrm{($\rho>0$, space-like extra dimension)},
\end{eqnarray}
\begin{eqnarray}
\label{neg}\eta=\pm\left.\int\sqrt{\left(\frac{\frac{8\pi
G}{3}|\rho| a^2}{1-\frac{8\pi G}{3}|\rho|
a^2}\right)}da\right|_{a=\sqrt{t'^2-r^2}},\\ \nonumber
\textrm{($\rho<0$, time-like extra dimension)}.
\end{eqnarray}
The energy density $\rho=\rho(a)$ is defined by the corresponding
equation(s) of state. Provided we know $\rho(a)$, we can easily
get particular equation for the submanifold. We can also see that
the choice (\ref{b2}) corresponds to the negative energy density
of the matter (in this case ${\dot a}^2<1$ as can be seen from
(\ref{Friedmann})).

Now we turn to the case of the closed universe. Let us take a
five-dimensional flat space-time with the metric
\begin{equation}\label{metric11}
ds^{2}=-dt'^{2}+d{\vec y}^{2}+
d\eta^2=-dt'^2+dr^2+r^2\left(d\theta^2+\sin^2(\theta)d\varphi^2\right)+
d\eta^2
\end{equation}
and make a coordinate transformation
\begin{eqnarray}\label{M11}
r=z\sin(\chi),\\ \label{M21} \eta=z\cos(\chi).
\end{eqnarray}
We obtain the metric
\begin{equation}
ds^2=-dt'^{2}+z^2\left(d\chi^2+\sin^2(\chi)\left(d\theta^2+\sin^2(\theta)d\varphi^2\right)\right)+dz^2.
\end{equation}
Now let us suppose that the coordinates on the hypersurface are
$\chi$, $\theta$, $\varphi$ and $t$, such that
\begin{equation}\label{M31}
z=a(t),\quad t'=b(t).
\end{equation}
We get
\begin{equation}\label{frw2}
ds^2=-\left({\dot b}^{2}(t)-{\dot
a}^{2}(t)\right)dt^2+a^{2}(t)\left(d\chi^2+\sin^2(\chi)\left(d\theta^2+\sin^2(\theta)d\varphi^2\right)\right).
\end{equation}
Thus if
\begin{eqnarray}\label{b11}
{\dot b}^{2}(t)={\dot a}^{2}(t)+1,
\end{eqnarray}
we get the four-dimensional FRW metric of the closed universe. We
see that now the only possibility for the ambient space-time is to
be (4+1), the case (3+2) being impossible. Equations (\ref{M11}),
(\ref{M21}), (\ref{M31}) and (\ref{b11}) can be found in
\cite{Robertson,Rosen,Robertson-Noonan,Lachieze-Rey}.

The equation of the appropriate submanifold can also be easily
obtained. Indeed, using (\ref{M11}), (\ref{M21}) and (\ref{M31})
we get
\begin{equation}\label{paramet2}
\eta^2+r^2=\left.a^2(t)\right|_{t=b^{-1}(t')}.
\end{equation}
To rewrite it in a more useful form we will repeat the steps
presented above for the open universe (but now using $\kappa=1$)
and obtain
\begin{equation}\label{clo_pos}
t'=\pm\left.\int\sqrt{\left(\frac{\frac{8\pi G}{3}\rho
a^2}{\frac{8\pi G}{3}\rho
a^2-1}\right)}da\right|_{a=\sqrt{\eta^2+r^2}}.
\end{equation}

For completeness we also present the formulas for embedding of the
spatially-flat FRW universe in five-dimensional Minkowski space:
\begin{eqnarray}\label{subst1}
t'&=&\frac{1}{\alpha}\left(a(t)\vec x^{2}+\int\frac{dt}{\dot a(t)}\right)+\alpha\frac{a(t)}{4},\\
\label{subst2} \eta&=&\gamma\left[\frac{1}{\alpha}\left(a(t)\vec
x^{2}+\int\frac{dt}{\dot a(t)}\right)-\alpha\frac{a(t)}{4}\right],\\
\label{subst3} \vec y&=&a(t)\vec x,
\end{eqnarray}
where $\gamma=\pm 1$, $\alpha$ is a constant with the dimension of
length, $\alpha\neq 0$. Substituting (\ref{subst1})-(\ref{subst3})
into (\ref{metric11}) we easily obtain
\begin{equation}\label{flatFRW}
ds^{2}=-dt^{2}+a^{2}(t)d{\vec x}^{2},
\end{equation}
which corresponds to a cosmology with zero spatial curvature. The
parameter $\alpha$ simply corresponds to the invariance of
(\ref{flatFRW}) under the rescaling $a(t)\to a(t)\beta$, $\vec
x\to \vec x/\beta$, where $\beta\ne 0$ is a constant. Equations
(\ref{subst1})-(\ref{subst3}) can be found in
\cite{Robertson,Robertson-Noonan} for arbitrary $\alpha$ and
$\gamma=-1$ (in our notations), in \cite{Rosen} with $\alpha=2$
and $\gamma=1$ and in \cite{Lachieze-Rey} with $\alpha=2$ and
$\gamma=-1$. They are also in agreement with the coordinate
transformations presented in \cite{Wessonetal,Wesson} between the
Ponce de Leon metric \cite{PdL} and the five-dimensional Minkowski
space (to see it one should set $\alpha\to 2\alpha$, $\gamma=-1$,
$a(t)=t^{\frac{1}{\alpha}}$ in (\ref{subst1})-(\ref{subst3}) and
$l\equiv 1$ in eqs. (4a)-(4c) of \cite{Wesson}).

Equation for the submanifold in five-dimensional Minkowski space
corresponding to the spatially-flat FRW universe has the form (it
can be obtained from the one presented in note D of
\cite{Robertson} (see also \cite{Smolyakov}) using the Friedmann
equation (\ref{Friedmann}) with $\kappa=0$):
\begin{equation}\label{manifold1rho}
t'^{2}-\vec y^{2}-\eta^{2}=\frac{3}{8\pi
G}\left[a\int\frac{da}{a^{2}\rho}\right]_{a=\frac{2(\gamma
t'-\eta)}{\alpha}}.
\end{equation}

Now let us turn to specific examples. Below we will present the
explicit form of the appropriate hypersurfaces for the
non-relativistic matter, the radiation and the cosmological
constant, all with $\rho>0$, i.e. for (4+1) ambient space-time. We
will omit the integration constants coming from (\ref{op_pos}),
(\ref{clo_pos}) and (\ref{manifold1rho}), they can be eliminated
by shifts of $t'$ or $\eta$. We also set $\gamma=1$ in
(\ref{manifold1rho}).

\begin{enumerate}
\item Matter dominated Universe,
$\rho=\rho_{0}a^{-3}.$
\begin{itemize}
\item Open universe.
$$\left(\frac{3}{16\pi G\rho_{0}}\eta\right)^{2}-\frac{3}{8\pi G\rho_{0}}\sqrt{t'^2-r^2}=1.$$
\item Closed universe.
$$\left(\frac{3}{16\pi G\rho_{0}}t'\right)^{2}+\frac{3}{8\pi G\rho_{0}}\sqrt{\eta^2+r^2}=1.$$
\item Spatially-flat universe.
$$
t'^{2}-r^{2}-\eta^{2}=C\left(t'-\eta\right)^{3},
$$
where $C\ne 0$ is arbitrary constant of the corresponding
dimensionality.
\end{itemize}
\item Radiation dominated Universe,
$\rho=\rho_{0}a^{-4}.$
\begin{itemize}
\item Open universe.
$$\sinh^2\left(\sqrt{\frac{3}{8\pi G\rho_{0}}}\,\eta\right)=\frac{3}{8\pi G\rho_{0}}\left(t'^2-r^2\right).$$
\item Closed universe.
$$\sin^2\left(\sqrt{\frac{3}{8\pi G\rho_{0}}}\,t'\right)=\frac{3}{8\pi G\rho_{0}}\left(\eta^2+r^2\right).$$
\item Spatially-flat universe.
$$
t'^{2}-r^{2}-\eta^{2}=C\left(t'-\eta\right)^{4}
$$
with $C\ne 0$.
\end{itemize}
\item Cosmological constant,
$\rho=\rho_{0}=\textrm{const}.$ For all three cases we get
$$
t'^2-r^2-\eta^2=-\frac{1}{H^2}
$$
with $H=\sqrt{\frac{8\pi G\rho_{0}}{3}}$, which is of course the
expected and the well-known result for $dS_{4}$ space.
\end{enumerate}
Note that since $a(t)$ in (\ref{frw1}) and (\ref{frw2}) has the
dimension of length, the dimension of $\rho_{0}$ in the above
formulas is different for different types of the matter.

Quite an interesting case is that of the cosmological constant
with $\rho=\rho_{0}<0.$ According to (\ref{neg}) we should take
(3+2) ambient space-time. From (\ref{neg}) we get the equation of
the hypersurface
$$
t'^2-r^2+\eta^2=\frac{3}{8\pi G|\rho_{0}|},
$$
which is the well-known result for $AdS_{4}$ space. We see that
for the negative energy density of matter the FRW metric exists
only for the case of the open universe and it can be represented
as the metric induced on the submanifold in (3+2) pseudo-Euclidean
space.

We would like to note that the simple examples presented above are
not the only equations of submanifolds which can be obtained
analytically. For example, for
$\rho(a)=\frac{\rho_{m}}{a^{3}}+\rho_{\Lambda}$ and $\kappa=0$,
which represents the main contribution to the energy density of
the Universe at the present epoch, the equation of the
corresponding submanifold can also be obtained analytically (but
this equation has a rather bulky form and we do not present it
here). As for more general cases, equations (\ref{op_pos}),
(\ref{neg}), (\ref{clo_pos}) and (\ref{manifold1rho}) allow one to
obtain analytically or numerically (and visualize, as it was made
in \cite{Lachieze-Rey,Wesson,Seahra,LB,Rindler2} for several
cases) the corresponding surfaces in five-dimensional
pseudo-Euclidean space using only equations of state of the matter
(of course, if these equations of state suggest an explicit form
of $\rho(a)$ without solving the Friedmann equations). It is worth
mentioning that such a visualization can be very informative, see
essay \cite{Wessonessay} on this subject.

One more remark is in order. The Friedmann equation
(\ref{Friedmann}) corresponds to the standard four-dimensional FRW
cosmology. In this sense the extra dimension in (\ref{metric}) is,
of course, unphysical. Meanwhile, there are many multidimensional
models which provide effective four-dimensional cosmology of the
FRW type, though with dynamics different from the conventional
one. For example, one can consider five-dimensional brane world
models \cite{branes} or five-dimensional Ricci-flat models
\cite{Wesson1,Wesson2}, where extra dimension is physical and the
five-dimensional space-time is not necessarily flat. In both cases
the effective four-dimensional FRW metric induced on the brane or
on some hypersurface in this physical five-dimensional space-time
again can be represented as the metric induced on the hypersurface
in the unphysical five-dimensional pseudo-Euclidean space with the
help of formulas (\ref{paramet1}), (\ref{paramet2}) and the
corresponding equation for the spatially-flat case (of course, we
can not use (\ref{op_pos}), (\ref{neg}), (\ref{clo_pos}) and
(\ref{manifold1rho}) because they were obtained with the help of
the standard Friedmann equation (\ref{Friedmann}), which is valid
only in the standard FRW cosmology). In the general case these
physical and unphysical five-dimensional space-times do not
coincide. But there are some exceptions. As examples, one can
consider the Ponce de Leon metric \cite{PdL} and the
five-dimensional analog of the de Sitter space \cite{PdL} (the
latter metric was also obtained in \cite{Liu}), these solutions
describe five-dimensional Ricci-flat space-times. It was noted in
\cite{Manus,Wesson3} that both solutions are Riemann-flat
space-times also. The explicit coordinate transformations between
the five-dimensional Minkowski space and these solutions can be
found in \cite{Wessonetal,Wesson} for the Ponce de Leon metric and
in, for example, \cite{Smolyakov2} for the five-dimensional analog
of the de Sitter space. Thus, in these cases the physical
five-dimensional space coincides with the unphysical one.

We hope that the result presented in this note can be useful, at
least from historical and pedagogical points of view.

\section*{Acknowledgments}

The authors are grateful to I.P.~Volobuev for valuable
discussions. M. S. acknowledges support of grant of Russian
Ministry of Education and Science NS-4142.2010.2, state contract
02.740.11.0244 and grant MK-3977.2011.2 of the President of
Russian Federation.

\end{document}